# Rocket Cratering in Simulated Lunar and Martian Environments

Christopher Immer[1], Philip Metzger[2]

[1]ASRC Aerospace, M/S ASRC-15, Kennendy Space Center, FL 32899; PH (321) 867-6752, FAX (321) 867-4766; email: christopher.immer@ksc.nasa.gov
[2]NASA/KSC Granular Mechanics and Surface Systems Laboratory, KT-D3, Kennedy Space Center, FL 32899; PH (321) 867-6052; FAX (321) 867-2975; email: Philip.T.Metzger@nasa.gov

## ABSTRACT

With NASA's planned return to the moon and possibly with lunar outposts being formed, repeated landings at the same site will be necessary. Understanding rocket plume interaction with lunar and Martian surfaces is of paramount importance in order to safely land and protect hardware surrounding the landing site. This work will report on results of three small experiments intended to explore plume impingement onto lunar and Martian surfaces: Handheld Observation of Scour Holes (HOOSH), Handheld Angle of Repose Measurements of Lunar Simulants (HARMLuS), and Mars Architecture Team study (MATS). The first two experiments were performed during two sorties of reduced gravity flights. HOOSH was designed to investigate crater formation as a function of gravitational level (lunar and Martian gravity). HARMLuS was designed to measure the Angle of Failure (related to the angle of repose) at lunar and Martian gravity. Both experiments have complex findings indicative of the hysteretic behavior of granular materials, especially resulting from reduced gravity. The MATS experiment was designed to investigate the effects of regolith compaction on the granular mechanics of crater formation. In general, the granular mechanics is a much stronger function of compaction than gravitation acceleration. Crater formation is greatly enhanced at reduced gravity (resulting in much larger craters). The angle of failure of the lunar simulants increases with decreasing gravitational acceleration, and occasionally becomes infinite for some compactions at lunar gravity. The angle of failure also increases with increasing compaction. While compaction does play a role in the time development of crater formation, the asymptotic behavior is largely unaffected.

## INTRODUCTION

The Surface Systems Group at Kennedy Space Center has been involved with many aspects of operating on celestial surfaces. Both the moon and Mars have a layer of regolith covering most of their surfaces.(Christensen 1986; Heiken, Vaniman and French 1991) The dusty, granular nature of these surfaces is challenging for operations.(Gaier 2005; Wagner 2006) The surface systems group has developed an extensive theoretical, experimental, and design program to aid in understanding and developing tools to property handle access to these types of regolith environments.(Donahue, Metzger and Immer 2006; Immer, Lane, Metzger et al. 2008; Metzger, Lane and Immer 2008; Metzger, Immer, Donahue et al. 2009;

Metzger, Latta, Schuler et al. 2009; Metzger, Lane, Immer et al. 2010) This paper will discuss three example experiments aimed at investigating crucial parameters for launch, landing, and operations on the moon and Mars. Those discussed here are simple experiments that are useful directly and are also designed to feed into theoretical models being developed. The combined theory/experiments will be used to design the architecture for missions to the moon and Mars.

The reduced/augmented gravity environments for the HOOSH and HARMLuS experiments were provided by NASA's Reduced Gravity Office at Johnson Space Center/Ellington Field, TX. They are obtained by flying an aircraft through parabolic trajectories. Reduced gravity environments are obtained at the peaks of the parabola and augmented gravity environments are obtained at the troughs of the flight pattern and through the turns at the ends of the airspace. Data were taken during stable acceleration periods of the parabolas. The approximate gravitational acceleration was obtained from wall mounted accelerometers aboard the aircraft.

# HOOSH

## Background

One common technique to investigate rocket exhaust cratering onto sandy surfaces is to split the symmetry of the exhaust nozzle with a beveled section of polycarbonate. The predominate erosion method for nozzle exhaust in this regime is *viscous erosion*. There have been numerous studies using this technique to investigate the parameters that effect cratering: exhaust velocity, nozzle diameter, gas density, grain size, grain density, etc…(Donahue, Metzger and Immer 2006; Metzger, Immer, Donahue et al. 2009; Metzger, Latta, Schuler et al. 2009) The Handheld Observation of Scour Holes experiment was designed to test the effect of gravitational acceleration on crater formation. This investigation is especially important for launch and landing on non-terrestrial bodies, such as the Moon and Mars that have lower gravitational acceleration than Earth.

## Experiment

Figure 1 shows a schematic of the HOOSH experiment. It consists of a 12x8" polycarbonate box containing the experiment. The box is double contained (not shown) to prevent leakage of the simulant. Inside the experiment is a well containing the simulant where one edge of the container has a raised beveled edge. An exit nozzle is centered above the beveled edge to split the exit flow and allow viewing of the crater formation. A filtered fan draws air from the front part of the chamber and directs it through the nozzle. A flow straightener in the upper part of the nozzle ensures that the flow from the nozzle is not turbulent. The fan can be run a different flow rates with an adjustable rheostat. 640x480 resolution video is recorded uncompressed at 8 Hz on a laptop with a firewire Prosilica camera. Video is recorded for approximately 30 seconds. After each run, the box is tilted backwards to place any ejected simulant back in the test bin. Figure 2 shows a picture of the HOOSH experiment on the reduced gravity flight.

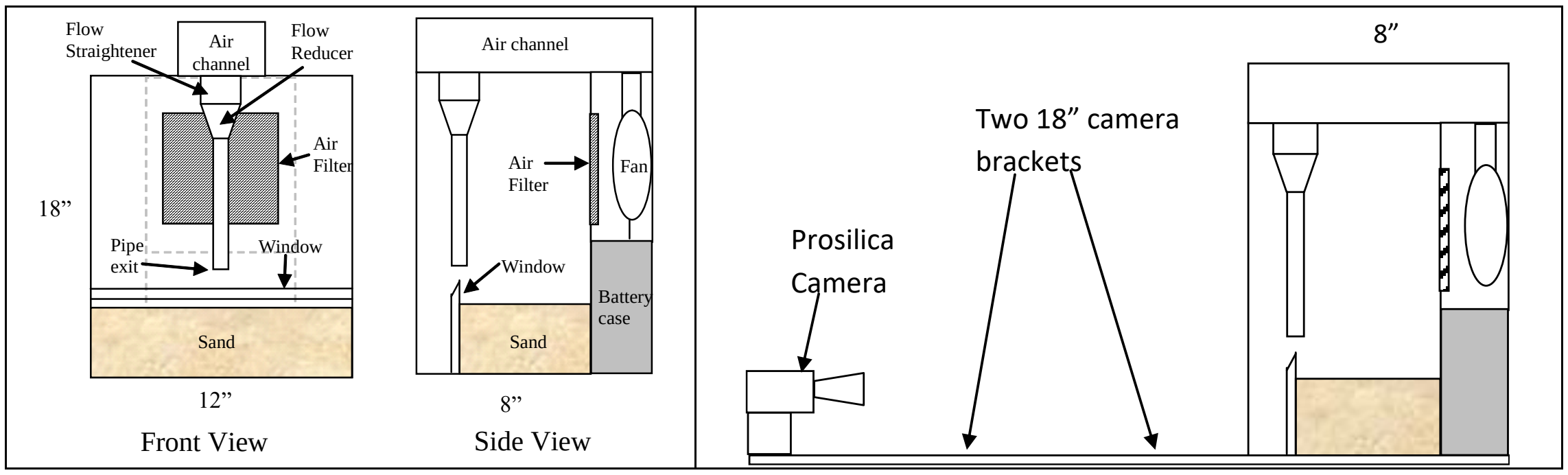

**Figure 1 HOOSH Experimental Setup**

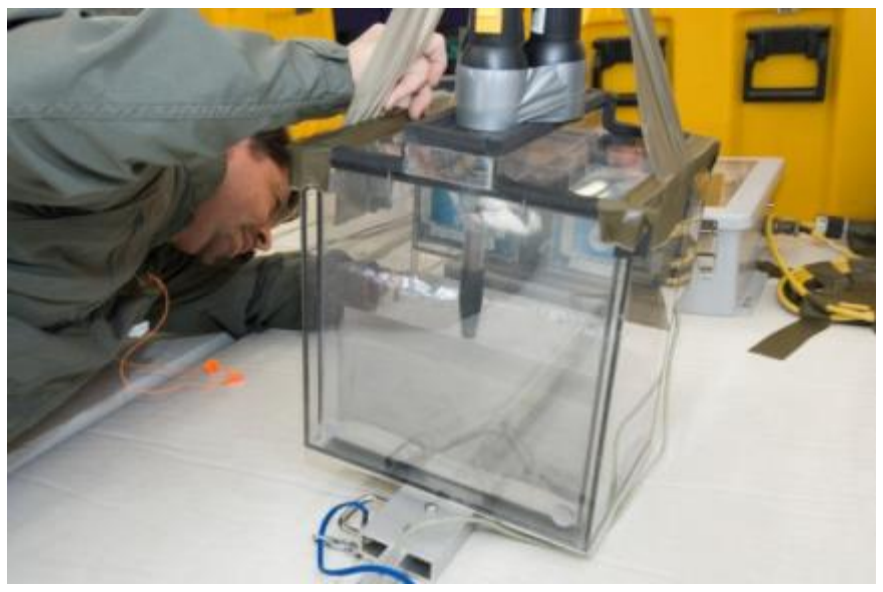

**Figure 2 Picture of HOOSH Experiment.**

Lighting directed vertically down onto the simulant bin causes a significant contrast between the crater and forward facing window. During post-processing, custom developed software is used that automatically measures the crater parameters for the recorded video. The software is able to extract depth, width, and an estimated crater volume vs. time from each of the videos in engineering units.

## Results and Discussion

The left half of Figure 3 shows a graph of the volumetric erosion rate vs. gravity and fan speed for Jetty Park beach sand sieved at 250-300 μm. These data were obtained by fitting the beginning linear portion of plots of crater volume vs. time for each of the experiments. Increasing fan speed number is increasing flow rate. For intermediate fan speeds, the erosion rate is inversely proportional to the gravitational acceleration. The right side of Figure 3 is a slice of the 3d graph at fan setting 8 showing the 1/g gravity dependence of the erosion rate. With lower flow rates the trend is almost linear in gravitational acceleration. For some of the higher fan speeds the erosion rate is low, probably due to turbulence in the flow path.

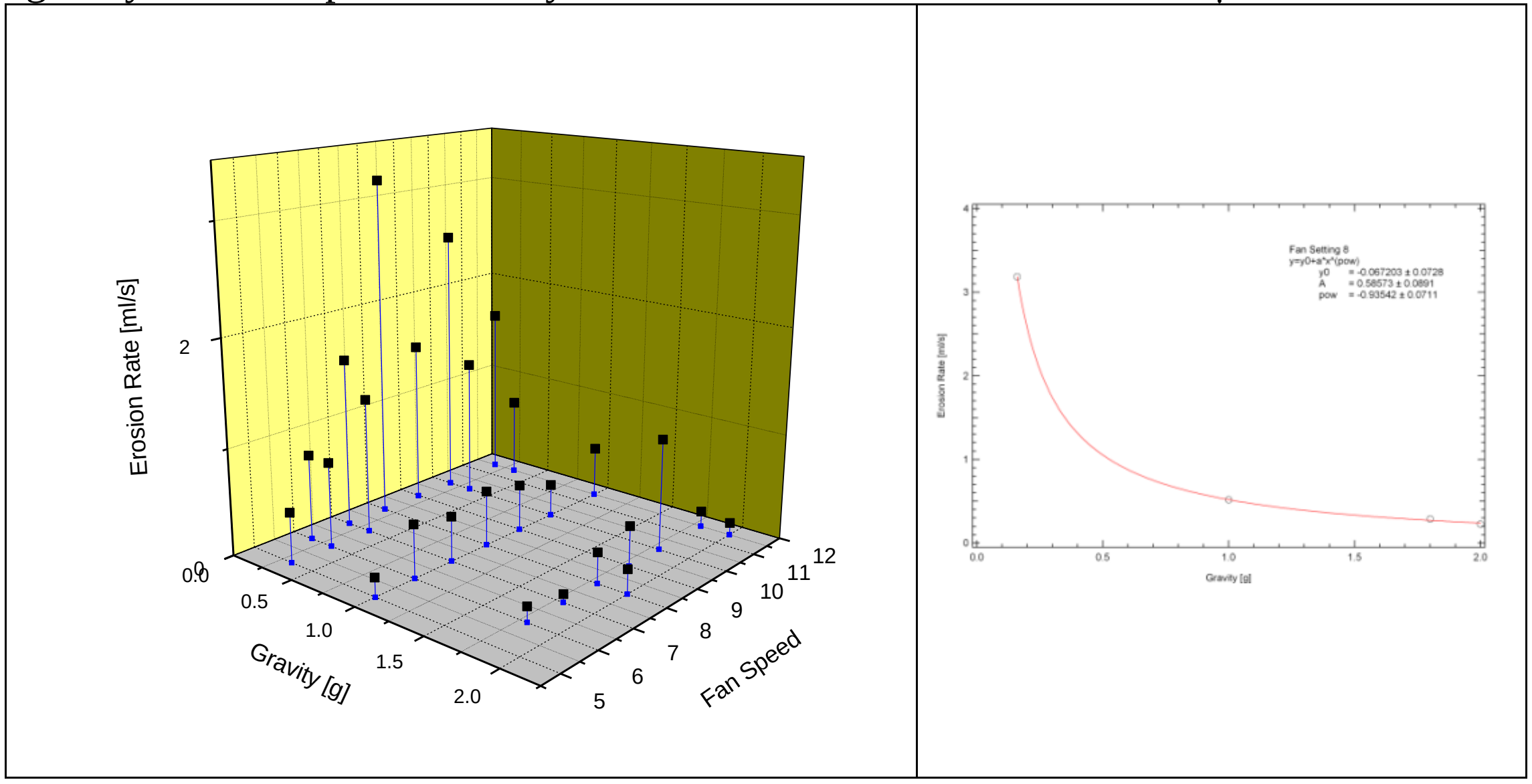

**Figure 3 HOOSH results showing volumetric Erosion Rate. The graph on the right is a slice of erosion rate vs. gravity for fan speed 8, showing the 1/g dependence.**

One of the complications with reduced gravity experiments and granular physics is that during the 1.8 g pull-outs the simulant is compacted. As can be seen with the HARMLuS experiments, the granular mechanics is a strong functions of the density: it’s difficult to ensure a consistent density. Since the true erosion rate is mass dependent, higher simulant densities will cause the volumetric erosion rate to appear slower. It is to be expected that some of the measurement uncertainty in the reduced gravity erosion rates is related to density variations.

## HARMLUS

### Background

Many regolith simulants have been developed to allow testing in terrestrial environments. One of the key parameters affecting the behavior of the simulants is the gravitational acceleration.(White and Klein 1990) In particular, the angle of repose (AOR) is an intrinsic property of a granular material that is the maximum angle of stable slope that is determined by gravity, friction, cohesion, and particle shape. Since typical experiments and testing occur in earth’s gravitational acceleration, it is important to know how the behavior of the simulants will change in Martian and Lunar gravity.

### Experiment

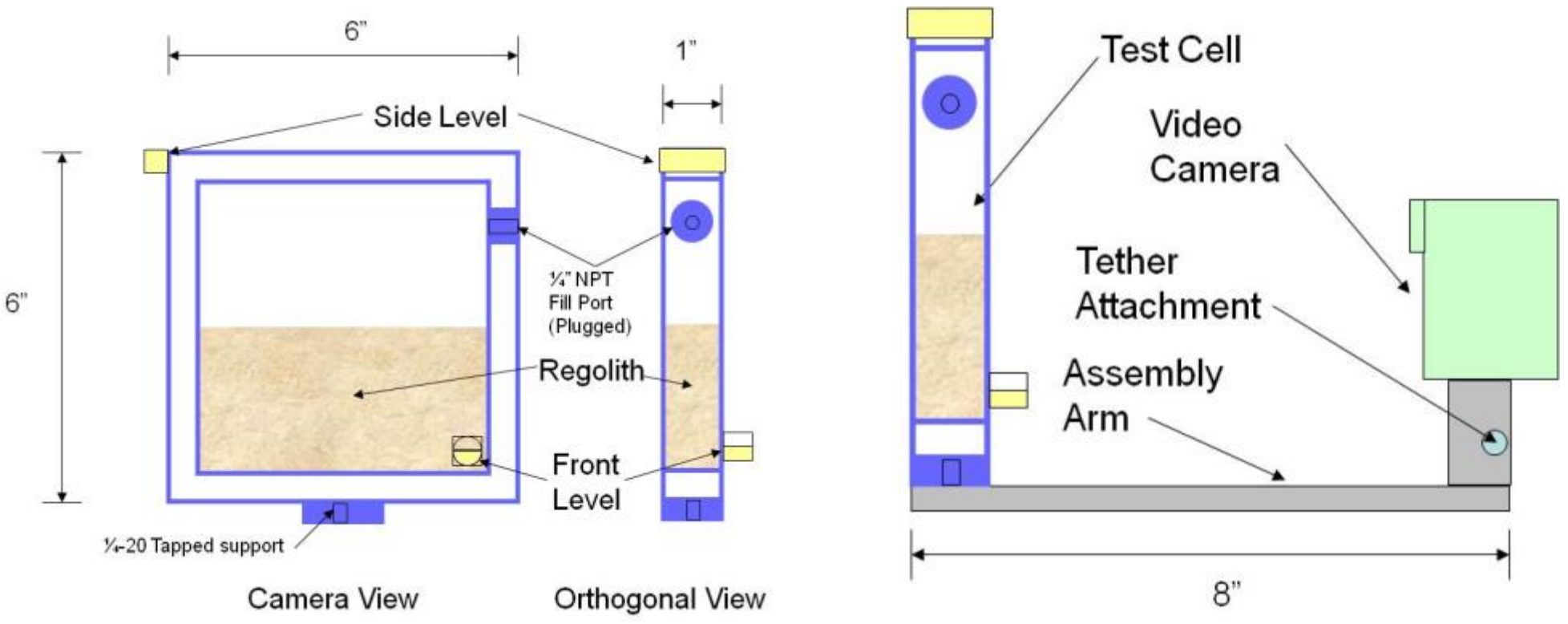

**Figure 4 HARMLuS Experiment.**

The intent of the HARMLuS was to measure the AOR of the granular materials, however the actual design of the experiment for the reduced gravity flight measures the Angle of Failure (AOF) of the material. While the AOR is measured by the angle formed by piling the granular material, the AOF is defined as the largest angle that a granular material can attain while being inclined before slip or failure. It is very closely related to the AOR and typically a few degrees steeper.

Figure 4 shows the design of the HARMLuS experiment. It consists of a set of four interchangeable polycarbonate cells (shown on the left), an assembly arm, a video camera, and a set of attached levels. Each of the four test cells are loaded with a known mass of simulant before the flight. The cells are at standard temperature and pressure with ambient humidity conditions.

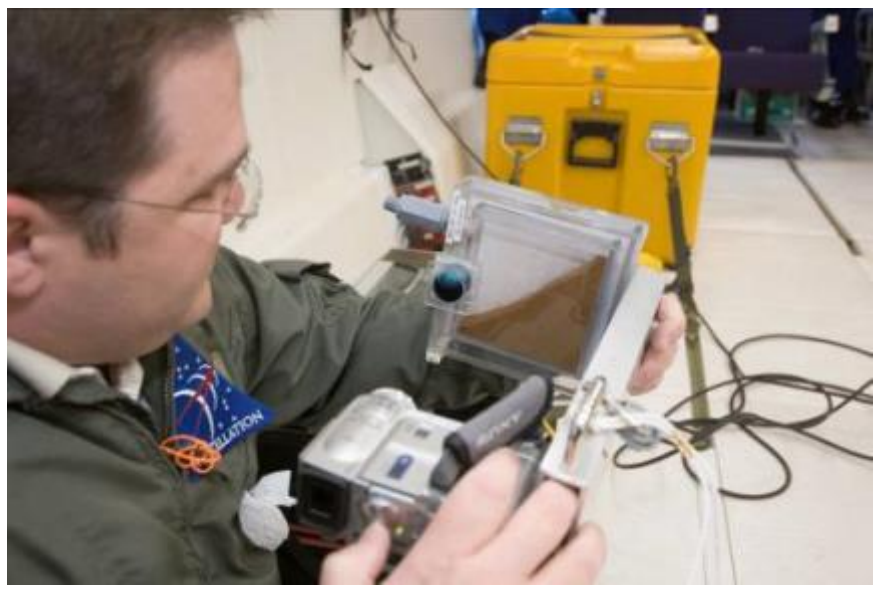

**Figure 5 HARMLuS Experiment.**

For operation of the experiment during periods of stable acceleration, the material in the cell is leveled, the video recorder is started, the parabola number and acceleration are announced, the cell is rotated around the assembly arm axis until the material in the cell slips, and then the recording is stopped. During acquisition of the data, every attempt is to keep the cell level with respect the side level. The front level is used to measure the effective net acceleration. This experiment was completed 1-2 times per stable acceleration period. Figure 5 shows a picture of the experiment after the simulant has "failed".

After the flights are complete the video data are analyzed with specially developed machine vision software. The software is able to extract the angle of failure with respect to the front bubble level (the effective net acceleration). The software is also able to extract the density of the simulant by measuring the volume of the material in the cell.

## Results and Discussion

Figure 6 shows the results of four different simulants for this experiment: Jetty Park Beach Sand, Lunar JSC-1a(McKay, Carter, Boles et al. 1993), OB-1, and NU-LHT-2M(Stoeser, Wilson, Weinstein et al. 2008). The Jetty Park Beach Sand is a good control group. The particles are rounded and exhibit low cohesion. There is a general trend of decreasing AOF with gravity, but there is no clear trend as a function of density.

Lunar JSC-1A results show stronger trends. With increasing gravitational acceleration, generally the AOF decreases. However the AOF is a much stronger function of density: with decreasing density, the AOF decreases, almost linearly.

OB-1 is a much more cohesive simulant. In general, the AOF increases with decreasing gravitational level and increasing density. The cohesion in the materials is strong enough at the high densities, that at 1/6g, the AOF is essentially infinite (shown as 90 degrees in the plots). This means that the cell can be turned completely upside down for these conditions without collapse of the material.

NU-LHT-2M is an extremely cohesive simulant. It displays the most hysteretic (for subsequent experiments at like-gravity) and ill-formed behavior. The trend is non monotic in either density or gravity. It displays some local miminum between the extrema. A high densities and low gravity it displays an infinite AOF.

In general for all four simulants studied, the AOF is a much stronger function of the material density than gravitational acceleration. As in the HOOSH study, one drawback of using an airplane for the reduced gravity environment for granular mechanics, is that the simulant get compacted during the 1.8g pull outs. It is difficult to separate the gravitational dependence from the density dependence. Uniformity of the density for these experiments plays a key roll in the uncertainty: it is difficult to get evenly distributed density throughout the sample.

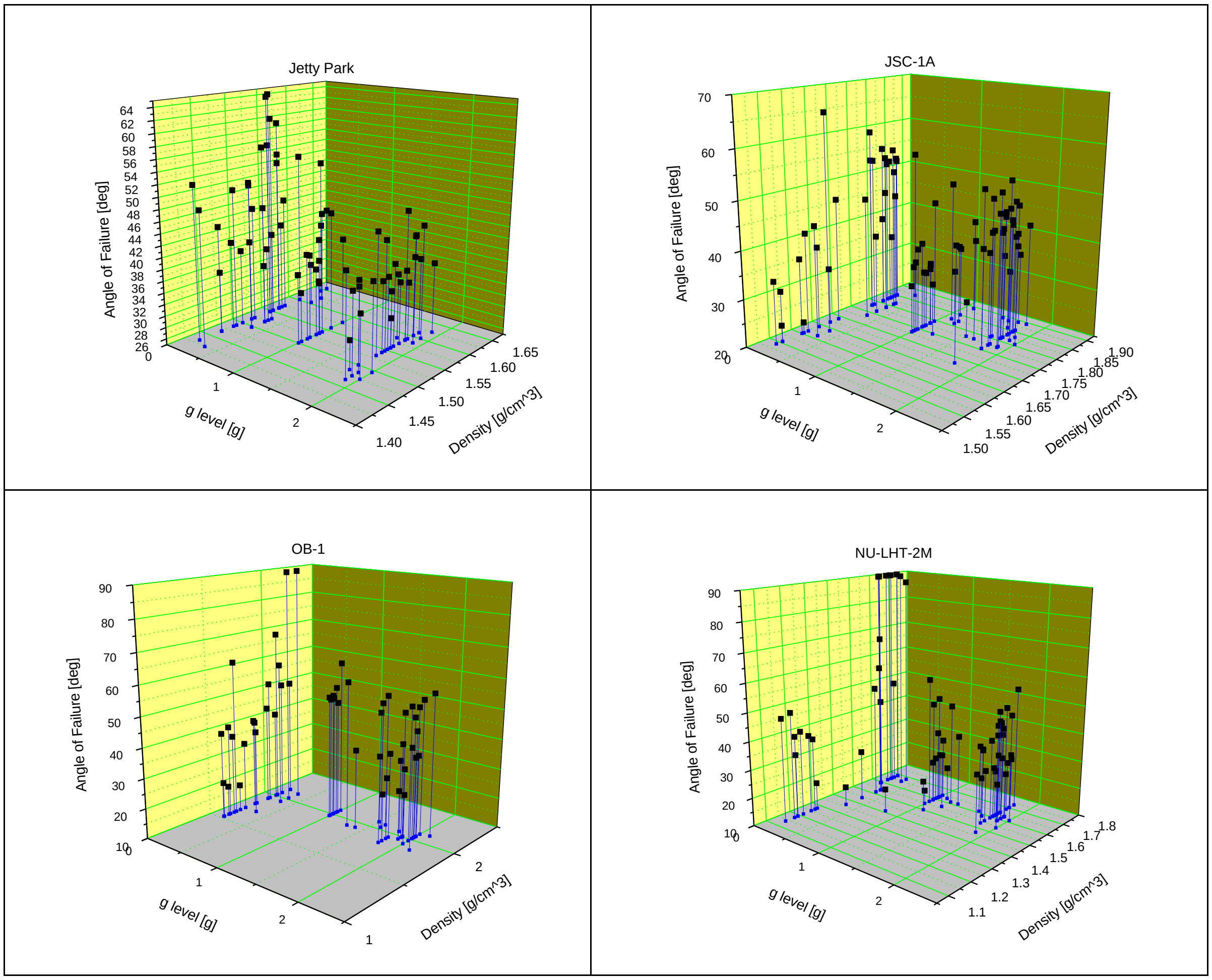

**Figure 6 HARMLuS results showing Jetty Park Beach Sand, Lunar JSC-1A, OB-1, and NU-LHT-2M**

# MATS

## Background

Human landers for the Martian surface will have much larger retrorockets than the unmanned missions have previously had. The Mars Architecture team approached our team to ask how these large retrorocket would affect the native regolith landing surface. Would the surface would be sufficient to support the lander? Is there anything that can be done to mitigate any blast effects?

While it is known that regolith density affects the crater formation, very little is known about how, in detail, it affects the formation. The density of regolith over the surface of Mars is expected to vary greatly from solid rock to loose sand, therefore, the selection of the landing site will affect the rocket plume crater formation. The lunar environment ranges from $10^{-12}$-$10^{-15}$ torr while the Martian environment is around 7 torr. While the Martian environment is substantially lower in ambient pressure, the presence of the small atmosphere is enough to collimate a rocket exhaust plume in a similar manner to the terrestrial environment. A key experiment would be to test regolith density dependence on crater formation, and performing the experiment in terrestrial atmosphere is sufficient to simulate the Martian environment.

## Experiment

The best simulant that we have, to date, is JSC Mars-1A, however it is not expected for this simulant to be high fidelity.(Allen, Jager, Morris et al. 1998) To match the conjectured particle size distribution of the Martian soil, ½ ton of 1 mm simulant was mixed with ½ ton of 5 mm simulant. The simulant was placed in a box approximately 0.3x1x3 $m^3$ with a front transparent polycarbonate window about 0.3x.0.3 $m^2$. A Vision Research Phantom 5.1, high speed camera was placed approximately 15 feet from the front polycarbonate window. High intensity light was used in place of solar lighting to mitigate shadowing. 1.2 seconds of video was acquired at 1200 frames per second for each of the experiments.

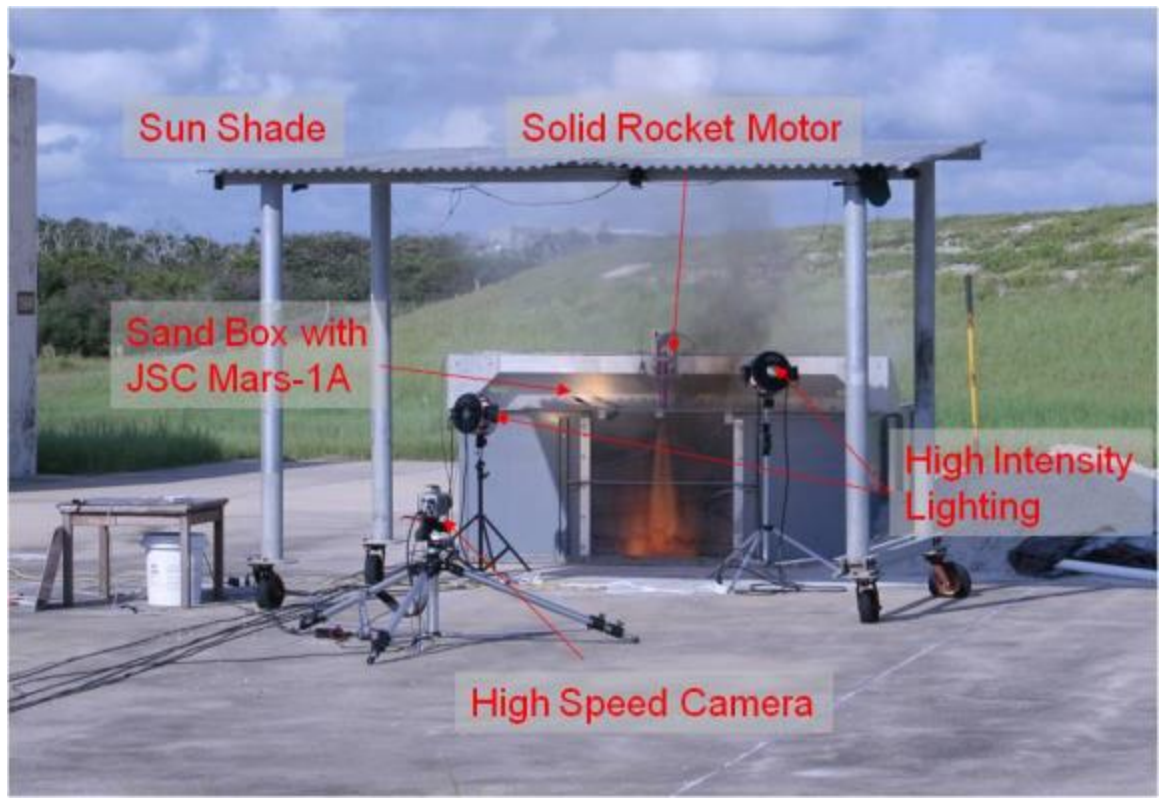

**Figure 7 MATS experiment setup.**

To create the crater, a solid rocket motor used was built in-house using HTPB/AP/Al propellant. The rocket plume was supersonic with an exit nozzle of about 1 inch and a thrust of about 370 N. The nozzle was placed a few centimeters above the surface of the simulant.

Compaction was performed by sequentially layering 3 cm layers of simulant and packing with a hand-held vibratory compaction device. The uncompacted sample was prepared by "fluffing" and stirring up the simulant. Handheld shearometer and penetrometers were used to measure the surface shear stress and cone penetration. The density was measured by massing a core of the simulant.

Three firings were completed with the rockets. Firing 1 was a test firing to ensure that the instrumentation and triggering were working properly. Firing 2 was compacted, and Firing 3 was uncompacted.

## Results and Discussion

**Table 1 Geotechnical Properties of MATS experiment simulant**

| | Firing 2, Compacted | Firing 3, Loose |
|---|---|---|
| Core Density [g/cm^3] | 1.04 | 0.25 |
| Shear Stress [kg/cm^2] | 2.20 | 0.70 |
| Pentration [kg/cm^2] | 9.70 | 1.40 |

Table 1 shows the geotechnical properties of the simulant for the two firings. The compacted density was about 4 times the uncompacted density.

The predominant means of erosion for this method of testing is expected to be *bearing capacity failure*. After firing on the compacted simulant, the shear strength of the simulant was high enough to support the side walls of the remnant crater. The remaining crater looked almost like a cored sample where the cylindrical walls of the core were stable from collapse. For the uncompacted crater, immediately after the motor thrust stopped, the sidewalls collapsed, resulting in a shallow conical shaped crater.

Figure 8 shows the depth of the crater vs. time for the compacted and uncompacted simulants that was extracted from the high speed video. The uncompacted crater develops approximately 3 times faster than the compacted crater. The asymptotic depth of the crater is almost unaffected by the density of the simulant and is roughly equivalent to the plume extinction length. This result indicates that the plume extinction length (vs soil mechanics) dictates the size of the crater.

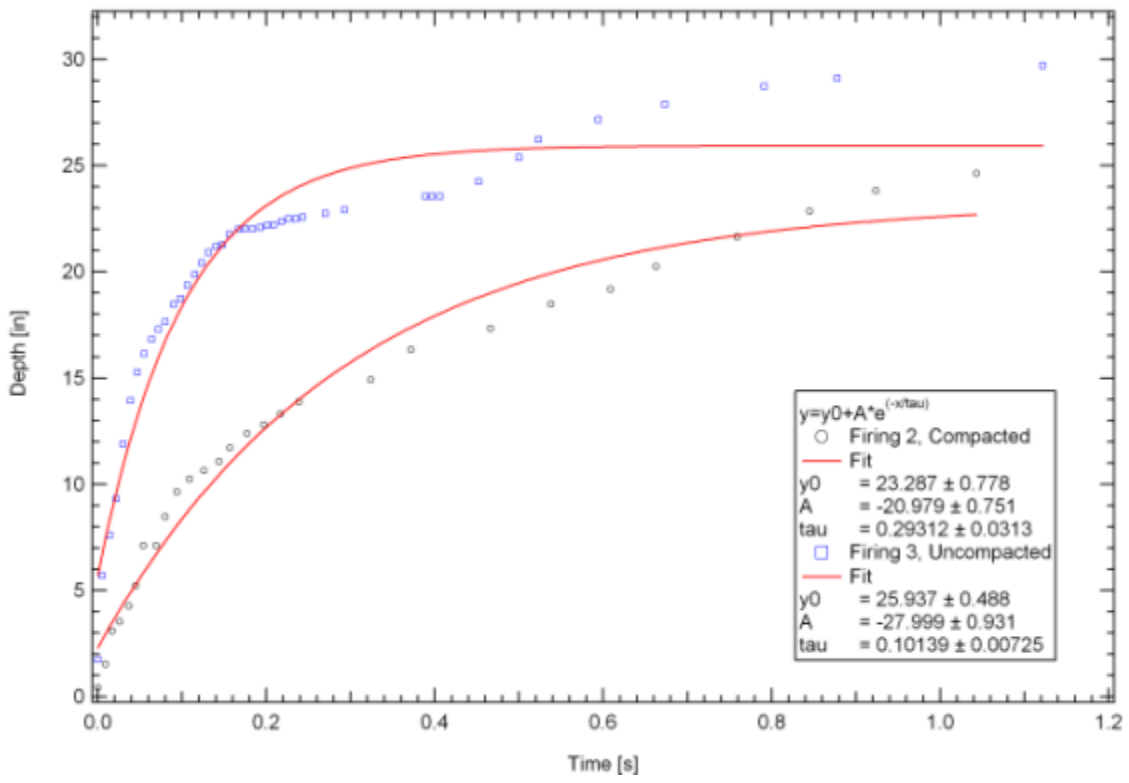

**Figure 8 Time evolution of the crater depth for compacted and uncompacted JSC-Mars-1a simulant.**

It has been estimated that for a 40 MT human sized lander, the plume extinction length could be anywhere from 6-9 m.(Metzger, Latta, Schuler et al. 2009) This would mean a crater would form beneath the lander on the order of 6-9 m. A crater of this size would certainly jeopardize the safety of the vehicle either for a remnant or collapsed crater. These results lead directly for the necessity of some mitigation strategy: stabilizing landing surface, redesigning the lander engine, not using retrorockets for landing, etc…

## CONCLUSION

These experiments have improved our understanding of the mechanics of lunar and martian soils. The granular mechanics are affected more by compaction than by gravitation acceleration. The angle of failure of the lunar simulants increases with increasing compaction and with decreasing gravitational acceleration, and occasionally becomes infinite for some compactions at lunar gravity. Crater formation under a subsonic gas jet is greatly enhanced at reduced gravity with a $1/g$ scaling. The asymptotic depth of craters formed under a supersonic gas jet is determined primarily by the length of the jet, but the rate of the crater's growth decreases with increasing soil compaction.